\documentstyle[11pt,newpasp,twoside,epsf]{article}
\def\edcomment#1{\iffalse\marginpar{\raggedright\sl#1\/}\else\relax\fi}
\reversemarginpar

\begin{document}
\title{AGB Stars as Tracers of Star Formation Histories: Implications
for GAIA Photometry and Spectroscopy}
 \author{A. Ku\v{c}inskas}
\affil{Lund Observatory, Box 43, SE-221 00 Lund, Sweden}
\affil{Institute of Theoretical Physics \& Astronomy, Go\v{s}tauto
12, Vilnius 2600, Lithuania}
\author{L. Lindegren}
\affil{Lund Observatory, Box 43, SE-221 00 Lund, Sweden}
\author{T. Tanab\'{e}}
\affil{Institute of Astronomy, The University of Tokyo, Tokyo,
181-015, Japan}
\author{V. Vansevi\v{c}ius}
\affil{Institute of Physics, Go\v{s}tauto 12, Vilnius 2600,
Lithuania}

\begin{abstract}
We argue that tracing star formation histories with GAIA using
main sequence turn-off (MSTO) point dating will mainly be
effective in cases of very mild interstellar extinction
($E_{B-V}<0.5$). For higher reddenings the MSTO approach will be
severely limited both in terms of traceable ages ($t<0.5$ Gyr at 8
kpc; $E_{B-V}=1.0$) and/or distances ($d=2$ kpc if $t\leq15$ Gyr;
$E_{B-V}=1.0$), since the MSTO will be located at magnitudes too
faint for GAIA. AGB stars may alternatively provide precise
population ages with GAIA for a wide range of ages and
metallicities, with traceable distances of up to 250 kpc at
$E_{B-V}=0$ (15 kpc if $E_{B-V}=2.0$). It is essential however
that effective temperatures, metallicities, and reddenings of
individual stars are derived with the precision of $\sigma(\log
T_{\rm eff})\sim0.01$, $\sigma([M/H])\sim0.2$, and
$\sigma(E_{B-V})\sim0.03$, to obtain $\sigma(\log t)\sim0.15$.
This task is quite challenging for GAIA photometry and
spectroscopy, though preliminary tests show that comparable
precisions may be achieved with GAIA medium band photometry.
\end{abstract}

\section{Introduction}

\begin{figure}[t]
\plotone{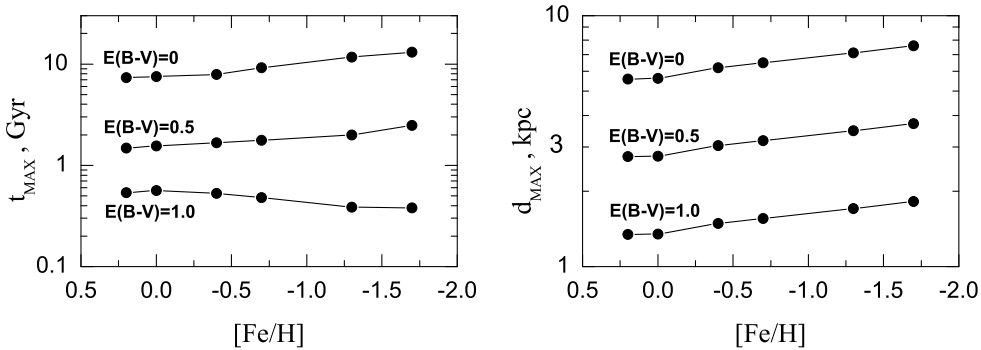}
\caption{Left: maximum age for which
MSTO point can be dated at 8 kpc with GAIA, for different [Fe/H]
and $E_{B-V}$. Right: maximum distance at which stellar population
up to 15 Gyr old can be dated using MSTO with GAIA, for different
[Fe/H] and $E_{B-V}$.}
\end{figure}

Out of the numerous methods available for assessing ages of
stellar populations only a few will be practically applicable for
use with GAIA, because of obvious limitations related to the
detection limits of GAIA etc.\ (e.g., Cacciari 2002). Though the
main sequence turn-off (MSTO) point approach is perhaps the most
reliable and accurate of all dating methods available today, it
will be severely distance-limited when used with GAIA.

The extent of these limitations is illustrated in Fig.~1. The left
panel shows the maximum age obtainable with GAIA employing MSTO
fitting for stellar population placed at a distance of 8 kpc
(assuming that the MSTO point is located at $V\sim18.5$, while the
detection limit of GAIA is $V=20$). The right panel shows the
maximum distance at which ages up to 15 Gyr may be quantified (in
both cases, the MSTO point on the isochrones of Girardi et al.\
(2000) was dated). It is obvious that in cases of negligible
interstellar extinction the MSTO approach may work very
efficiently with GAIA, providing reliable ages of stellar
populations up to $\sim8$ kpc. However, even mild interstellar
extinction changes the situation dramatically, shifting the MSTO
point below the detection limits of GAIA for many stellar
populations within the Galaxy.

\section{AGB stars with GAIA: star formation histories}

\begin{figure}[t]
\plotone{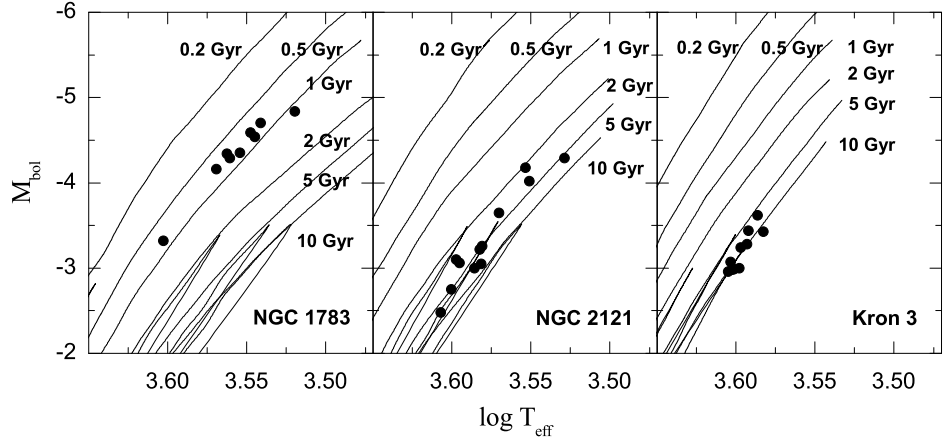}
\caption{Observed HR diagrams with
AGB sequences in star clusters of the Magellanic Clouds: NGC 1783,
2121 (both LMC) and Kron 3 (SMC). Isochrones are from Bertelli et
al. (1994; B94).}
\end{figure}

A possible way to complement the MSTO approach would be with
tracers that are intrinsically brighter than MSTO stars, which could
thus be used for probing stellar populations on larger distance
scales. We have shown, that stars on early-AGB (non-thermally
pulsing) may be well suited for this purpose, providing reliable
ages of stellar populations from the isochrone fitting to AGB
sequences on the observed HR diagrams (Ku\v{c}inskas et al. 2002).

Though similar approaches have been used in the past (e.g.,
employing AGB sequences in color-magnitude diagrams), the
precision in derived ages was low. We propose to work in the
$M_{\rm bol}$ vs.\ $T_{\rm eff}$ plane, using effective
temperatures of individual AGB stars derived from the fits of the
observed spectral energy distributions (SEDs) with synthetic SEDs.
This yields a significant increase in precision of derived
population ages, which is a result of the precisely derived
$T_{\rm eff}$ of individual AGB stars. Since distances will be
known for the majority of Galactic AGB stars with GAIA (which,
combined with data from ground- and/or space-based infrared
surveys, will give precise $M_{\rm bol}$), AGB stars will allow
stellar populations on large distance scales to be traced with
GAIA.

\begin{table}[b]
\caption{AGB ages of star clusters in the Magellanic Clouds
derived in this work using different sets of isochrones (two
metallicities are used for Kron 3).}
\begin{tabular}{ccccccc}
\\
\tableline
\noalign{\smallskip}
Cluster     &     [Fe/H]    & \multicolumn{3}{c}{AGB ages, this work (Gyr)} &   MSTO ages   &  Ref  \\
            &               &      B94      &      G00      &      LS01     &               &       \\
\noalign{\smallskip}
\tableline
\noalign{\smallskip}
NGC 1783    &     $-0.4$    &  $0.8\pm0.1$  &  $1.1\pm0.2$  &       --      &  $0.9\pm0.4$  &  M89  \\
NGC 2121    &     $-0.7$    &  $3.8\pm0.8$  &  $6.0\pm1.6$  &       --      &  $3.2\pm0.5$  &  R01  \\
Kron 3      &     $-1.3$    & $10.3\pm2.8~$ &     $>15$     &  $7.0\pm1.6$  &  $8.0\pm0.5$  &  R00  \\
            &     $-1.0$    &  $4.9\pm1.1$  &     $>15$     &       --      &  $6.0\pm1.3$  &  M98  \\
\noalign{\smallskip}
\tableline
\end{tabular}
\end{table}

We illustrate the proposed approach employing the star clusters
NGC 1783, NGC 2121 and Kron 3 in the Magellanic Clouds. The
observed SEDs of AGB stars were constructed using {\it BVRIJHK\/}
fluxes obtained from the literature. The BaSel~2.2 library of
stellar spectra (Lejeune et al.\ 1998) was used to produce a
template of synthetic photometric colors at different $T_{\rm
eff}$ and for the metallicities of individual MC clusters. HR
diagrams of individual clusters with AGB sequences were
constructed using $M_{\rm bol}$ derived from observed integrated
fluxes of individual stars and $T_{\rm eff}$ obtained from the SED
fitting procedure (Fig.~2). AGB ages of individual clusters
derived using different sets of isochrones, as well as those
obtained from MSTO in previous studies, are given in Table 1.

There is indeed a good agreement between AGB ages obtained using
iso\-chrones of Bertelli et al.\ (1994; B94) or Lejeune \&
Schaerer (2001; LS01) on one hand, and MSTO estimates on the
other. AGB ages obtained with Girardi et al.\ (2000; G00)
isochrones are however considerably older than MSTO estimates
(Table~1). This effect tends to increase with decreasing
metallicity. It should be stressed, thus, that these intrinsic
discrepancies between different sets of isochrones clearly
indicate a need for better understanding of stellar evolution on
the AGB, to fully exploit the potential of AGB stars for tracing
stellar populations in the Galaxy and beyond.

Possibilities for tracing star formation histories using AGB stars
with GAIA, employing photometric metallicities, gravities and
reddenings (GAIA 1X medium band photometric system,
Vansevi\v{c}ius \& Brid\v{z}ius 2002) are summarized in Table~2
(the error in age for a single star, $\sigma(\log t)$, is a lower
limit since it reflects only the errors in $T_{\rm eff}$ and
$E_{B-V}$). We conclude that, contrary to MSTO dating, AGB stars
may provide precise estimates of ages ($t>0.5$ Gyr) throughout the
Galaxy and beyond ($d\la250$ kpc) for populations within a large
range of metallicities, even if interstellar extinction is
non-negligible. To achieve this, metallicities to $\pm0.2$~dex,
and effective temperatures to $\pm0.01$~dex are highly desirable.
A precise knowledge of interstellar reddening
($\sigma(E_{B-V})\sim0.03$) is also essential.

\begin{table}[t]
\caption{Predicted accuracies, versus $V$ magnitude, for $T_{\rm
eff}$, [M/H], $E_{B-V}$ (individual AGB stars, 1X photometric
system -- Vansevi\v{c}ius \& Brid\v{z}ius 2002), and ages from AGB
stars derived with GAIA. $d$ is the maximum distance at which
these accuracies may be expected.}
\begin{tabular}{cccccccc}
\\
\tableline \noalign{\smallskip}
$V$   & $\sigma(\log T_{\rm eff})$&  $\sigma([M/H])$  &  $\sigma(E_{B-V})$  &  $\sigma(\log t)$   &  $d_{E_{B-V}=0}$  &  $d_{E_{B-V}=2}$ \\
      &                           &                   &                     &                     &        kpc        &       kpc        \\
\noalign{\smallskip}
\tableline
\noalign{\smallskip}
18    &            $<0.01~$       &            $<0.2~$&           $<0.03~$  &         0.15        &        100        & \phantom{0}6     \\
20    &  $\phantom{<}0.03$        &  $\phantom{<}0.3$ & $\phantom{<}0.06$   &         0.45        &        250        &           15     \\
\noalign{\smallskip}
\tableline
\end{tabular}
\end{table}

\acknowledgments

This work was supported by the Wenner-Gren Foundations.



\begin{references}

\reference Bertelli, G., Bressan, A., Chiosi, C., Fagotto, F., Nasi, E. 1994, A\&AS, 106, 275 (B94)
\reference Cacciari, C. 2002, EAS Publ. Ser., 2, 163
\reference Girardi, L., Bressan, A., Bertelli, G., Chiosi, C. 2000, A\&AS, 141, 371 (G00)
\reference Ku\v{c}inskas, A., Vansevi\v{c}ius, V., Tanab\'{e}, T. 2002, Ap\&SS, 280,151
\reference Lejeune, T., Schaerer, D. 2001, A\&A, 366, 538 (LS01)
\reference Lejeune, T., Cuisinier, F., Buser, R. 1998, A\&AS, 130, 65
\reference Marigo, P., Girardi, L. 2001, A\&A, 377, 132
\reference Mighell, K.J., Sarajedini, A., French, R.S. 1998, AJ, 116, 2395 (M98)
\reference Rich, R.M., Shara, M.M., Fall, S.M., Zurek, D. 2000, AJ, 119, 197 (R00)
\reference Rich, R.M., Shara, M.M., Zurek, D. 2001, AJ, 122, 842 (R01)
\reference Mould, J., Kristian, J., Nemec, J., Jensen, J., Aaronson, M. 1989 ApJ, 339, 84 (M89)
\reference Vansevi\v{c}ius, V., Brid\v{z}ius, A. 2002, this volume

\end{references}
\end{document}